\documentclass{article}
\pdfoutput=1
\usepackage{microtype}
\usepackage{graphicx}
\usepackage{subfigure}
\usepackage{booktabs} 

\usepackage[accepted, nohyperref]{icml2021}

\icmltitlerunning{Driving User Satisfaction via Reward Shaping in a REINFORCE Recommender}

\begin{document}

\twocolumn[
\icmltitle{Reward Shaping for User Satisfaction in a REINFORCE Recommender}




\begin{icmlauthorlist}
\icmlauthor{Konstantina Christakopoulou}{goo}
\icmlauthor{Can Xu}{goo}
\icmlauthor{Sai Zhang}{goo}
\icmlauthor{Sriraj Badam}{goo}
\icmlauthor{Trevor Potter}{goo}
\icmlauthor{Daniel Li}{goo}
\icmlauthor{Hao Wan}{goo}
\icmlauthor{Xinyang Yi}{goo}
\icmlauthor{Ya Le}{goo}
\icmlauthor{Chris Berg}{goo}
\icmlauthor{Eric Bencomo Dixon}{goo}
\icmlauthor{Ed H. Chi}{goo}
\icmlauthor{Minmin Chen}{goo}
\end{icmlauthorlist}

\icmlaffiliation{goo}{Google, Inc, Mountain View, CA, United States}

\icmlcorrespondingauthor{Konstantina Christakopoulou}{konchris@google.com}

\icmlkeywords{Reward Shaping, User Satisfaction, Recommender  System}

\vskip 0.3in
]



\printAffiliationsAndNotice{}  

\begin{abstract}
How might we design Reinforcement Learning (RL)-based recommenders that encourage aligning user trajectories with the underlying user satisfaction? Three research questions are key: (1) measuring user satisfaction, (2) combatting sparsity of satisfaction signals, and (3) adapting the training of the recommender agent to maximize satisfaction. 
For measurement, it has been found that surveys explicitly asking users to rate their experience with consumed items can provide valuable orthogonal information to the engagement/interaction data, acting as a proxy to the underlying user satisfaction.
For sparsity, i.e, only being able to observe how satisfied users are with a tiny fraction of user-item interactions, \emph{imputation models} can be useful in predicting satisfaction level for all items users have consumed. For learning satisfying recommender policies, we postulate that reward shaping in RL recommender agents is powerful for driving satisfying user experiences. Putting everything together, we propose to jointly learn a policy network and a satisfaction imputation network: The role of the imputation network is to learn which actions are satisfying to the user; while the policy network, built on top of REINFORCE, decides which items to recommend, with the reward utilizing the imputed satisfaction. We use both offline analysis and live experiments in an industrial large-scale recommendation platform to demonstrate the promise of our approach for satisfying user experiences. 
\end{abstract}


\section{Introduction}
\label{sec:intro}
Recommender systems at heart aim at creating a good user experience by surfacing users with the right content at the right time and under the right context. 
It is thus critical for the system to identify what defines the user experience, more specifically the underlying user utilities of the platform. Most recent advances in recommender systems have relied on implicit user feedback, such as clicks or dwell time, as proxies to capture user utilities \cite{covington2016deep, yi2014beyond}.  
Although this data measures \emph{what users do}, it can fail to capture what users \emph{say they want} --- which are potentially very different \cite{lalmas2019reveal}. 
As a result, recommender models learned solely based on user engagement data can be misaligned with the true user utilities.  
 Given a specified objective, or else \emph{reward function}, which captures the long-term user utility, recommender systems can be formulated as Reinforcement Learning  (RL) agents deciding on actions to take (i.e., contents to show to users) given certain user states (i.e., latent representation at a specific time/context), with the goal of maximizing said cumulative reward~\cite{chen2019top,liu2018deep,48200,zhao2018deep}. There are several challenges especially exacerbated in industrial recommendation settings which makes the application of RL for recommendation rather unique compared to other application areas like games \cite{mnih2015human, silver2018general} and robotics \cite{gu2017deep}.
The action space is extremely large and  ever-changing; user preferences change over time; and data are extremely sparse for the enormous action and state space. 
Only recently there have been major advances addressing these challenges and showcasing RL approaches for recommendation \cite{chen2019top, liu2018deep}.
 
 
Besides addressing these challenges, the key in building recommender agents lies in defining the reward function guiding the learning of the agent policy. 
 Although we do get to observe some proxy signals indicating when a recommendation, or a series of recommendations was successful (e.g., the user clicked on the recommended content, they shared it with their social network etc.), there is a disconnect between the implicit feedback we observe, and what the  user really  wants. 
 The proximity between the proxy signals we include in the agent's reward function and the true user utility, will largely determine the extent to which the RL recommender can optimize for what users want.

However, despite the importance of the reward function for building recommendation agents, there has been relatively little work on reward shaping for RL recommenders. Most works treat the reward function as a black-box, which is given, and often assume that dense engagement signals are indicative of how much users value their experience. This assumption has been recently challenged in non-RL settings, underlying that post-engagement signals and/or satisfaction survey responses together with implicit behavioral signals give a clearer picture of user utilities \cite{guo2012beyond, lalmas2019reveal, wen2019leveraging}. 

In this paper, we put the reward front and center, and highlight it as a key tool for optimizing for what users actually want. 
Satisfaction data as collected by user responses to satisfaction surveys provide an important view as to how the user felt about the recommendation, as opposed to how they behaved while interacting with it. These surveys are shown uniformly to all users, and ask users to rate on a scale how satisfying they found a sampled item from
their recent engagement history. In our systems, such survey data can offer more representativeness compared to post-engagement signals, as most users tend to not engage in post-click actions such as likes, dismissals. Furthermore, optimizing for satisfaction as measured by surveys can  substantially move post-engagement related metrics as well. Based on the above, both behavioral signals, and satisfaction signals should be incorporated into the reward. 

However, before we can utilize satisfaction signals into the reward, we need to highlight a major  challenge inherently associated with them --- \emph{sparsity}. The volume of satisfaction data is orders of magnitude smaller compared with engagement data volume; in our case study, roughly \emph{one out of thousands} of engagement signals will come with a satisfaction response. This is due to a number of reasons. First, it is disruptive to ask users about every item they recently consumed. Second, response rate can be very low in an environment where primary user intention is to consume content rather than providing feedback. 
As a result, we only have access to a small amount of survey responses covering an extremely small fraction of the user-item interaction pairs. 
Given this extreme sparsity, simply supplementing the existing reward signals that focus on engagement with the sparse satisfaction signals is not going to be effective in shifting towards optimizing for user satisfaction. 
Instead, personalized satisfaction models are required to impute for each user how they would rate their satisfaction level with each consumed item, had they responded to a survey.

Here, we propose 
 augmenting a classic policy network trained with REINFORCE with a satisfaction imputation network to predict user satisfaction and include the prediction into the reward for the policy network, while training both networks concurrently in a multi-task learning setup. 
Together we offer the following contributions: 
\begin{itemize}
    \item \textbf{Reward Shaping as a tool to align with user utilities:} We highlight reward shaping as a tool to guide RL-based recommender agents in selecting satisfying actions for the users (Sections \ref{subsec:reward}, \ref{subsec:satisfaction-data}, \ref{sec:live-experiments}), which has been largely overlooked in advances on RL recommenders. In the process, we draw attention to common challenges associated with defining, measuring, and modeling user satisfaction.
    \item \textbf{ Satisfaction Imputation networks at scale:} We offer satisfaction imputation networks to combat sparsity of available utility signals capturing satisfaction, and demonstrate their usage to a top-$K$ REINFORCE recommender with an extremely large state and action space (Section \ref{subsec:actor-critic}).  We include offline analysis providing insights on design choices important when building a good satisfaction imputation network (Section \ref{sec:critic-results}).
    \item \textbf{Benefits in Live Experiments:} We provide evidence from A/B experiments in a large-scale recommendation platform with a two-stage nomination-ranking recommendation setup, that when we replace a REINFORCE nominator with dense engagement-based reward and  without a satisfaction imputation network with our  proposed architecture (Figure \ref{fig:actor-critic}), the number of satisfying nominations increases, while nominations with low satisfaction score decrease. This leads to a statistically significant improvement in satisfying user experiences, and decrease in unsatisfying experiences
    (Section \ref{sec:live-experiments}).
\end{itemize}


\section{Related Work}
\label{sec:related-work}
Here, we give an overview of the most closely related works.

  \textbf{Reinforcement Learning (RL).} Problems in which an agent learns to interact with the environment, 
  with the interactions having long-term consequences,  are
  a natural fit to be framed as Reinforcement Learning ones \cite{sutton2018reinforcement}. 
Classical approaches to RL problems include value-based approaches such as Q-learning \cite{mnih2013playing},
and policy-based ones such as policy gradient \cite{williams1992simple}.
Deep RL combines the promise of deep neural networks to help RL achieve ground-breaking success in games and robotics applications~\cite{mnih2015human, mnih2016asynchronous, silver2018general, gu2017deep}. 
 We build our work on top of a policy-based approach, namely REINFORCE \cite{williams1992simple}, following its prior success in recommendation settings~\cite{chen2019top}.
 The imputation network we introduce has deep connections with value learning approaches \cite{mnih2013playing}, where a state-action value network is learned. 
 The estimations of this network are utilized as \emph{part} of our policy's reward; 
as a result, we still need the off-policy correction component \cite{chen2019top}. An alternative approach we leave for future work is employing Actor-Critic or its variants 
 \cite{mnih2016asynchronous,sutton2000policy,schulman2015high}.

  \textbf{Reinforcement Learning in Recommendation.} Although there have been many successes in RL for applications like games \cite{mnih2015human, silver2018general} and robotics \cite{gu2017deep}, only recently some successes of RL for recommendation have been demonstrated \cite{dulac2015deep}. 
The main work we build on top of is a policy-gradient-based approach correcting for off-policy skew with importance weighting \cite{chen2019top}.
This work demonstrated the value of REINFORCE and top-$K$ off-policy correction in a large-scale industrial recommendation platform with an extremely large action space. Other recent works have demonstrated the value of deep RL approaches for recommendation, such as Actor-Critic \cite{liu2018deep}, Deep Q-learning \cite{zheng2018drn}, and hierarchical RL \cite{zhang2019hierarchical}. Also, novel RL approaches have been proposed for the more complicated problem of slate recommendation \cite{48200}, as well as for page-wise recommendation \cite{zhao2018deep}. Despite the recent promise of RL for recommendation, the majority of works do not draw attention to the important aspect of reward shaping, which is key for aligning system objectives with underlying user utilities; this is the focus of our paper.

 \emph{Reward Shaping.} 
The importance of reward shaping, i.e., shaping the original sparse, delayed reward signals as in-time credit assignment for successful RL algorithms has been emphasized early on \cite{ng1999policy, mataric1994reward, dorigo1994robot}. This is a general term encompassing the incorporation of domain knowledge into RL to guide the policy learning. 
Carefully designing the reward function is critically important as: (i) a misspecified reward leads to sub-optimal policy; 
(ii) an under-specified reward leads to unexpected behavior~\cite{hadfield2017inverse}. 
While RL for robotics and games has relied on hand-crafted reward or imitation learning~\cite{schaal1999imitation} to effectively guide the agent to success, the perils of a misspecified reward function in the design of recommender RL agents have not received a lot of attention. Motivated by the need to bridge the RL for recommendation line of work with the recent discussions on measuring and modeling user satisfaction to properly capture user experience  \cite{mehrotra2019jointly,lalmas2019reveal,mehrotra2018towards,garcia2018understanding}, we provide a reward shaping approach for imputing user satisfaction into the reward of an RL recommender, along with using the ground truth engagement proxy signals.     


\section{Background}
\label{sec:background}
We start by introducing useful background on RL for Recommendation, and notations.

\subsection{Recommendation as an RL problem}
The recommender system's goal is to decide which contents to recommend to the incoming user requests, given some representation of the user profile, the context, and their interaction history up to this point, as captured by the sequence of items (e.g. videos, news articles, products) they have interacted with, along with the corresponding feedback (e.g., time spent watching/reading), so as to maximize the cumulative rewards experienced by the users.

In RL terms, we formulate the recommendation problem as a Markov Decision Process (MDP)~\footnote{It is in fact a Partially Observed MDP (POMDP) \cite{sutton2018reinforcement} as the states are not fully observed.}: 

\begin{tabular}{|p{0.15\textwidth}p{0.25\textwidth}|}
\hline
\multicolumn{2}{|c|}{\textbf{Recommendation MDP}} \\
\hline
\textbf{Action}~ $a \in \mathcal{A}$  & item(s) available for recommendation \\
\textbf{State}~ $s \in \mathcal{S}$  & user interests and context \\
\textbf{State Transition} ${s}_{t+1} \sim \textbf{P}(\cdot | {s}_t, a_t)$ & unknown dynamics capturing how user state changes from $t$ to $t+1$, conditioned on $a_{t}$ and ${s}_t$ \\
\textbf{Reward}~ $r({s}, a)$ & immediate reward obtained by performing action $a$ for state ${s}$\\
\hline
\end{tabular}


The goal is to find a policy $\pi(a|{s})$ capturing the probability distribution over the action space, i.e, items to recommend, given the current user state ${s} \in \mathcal{S}$, so to maximize the expected cumulative reward,
\begin{equation}
\label{eq:1}
    \max_{\pi} \mathbf{E}_{\tau \sim \pi} \left[R(\tau)\right] 
\end{equation} where $ R(\tau) = \sum_{t=0}^{|\tau|} r({s}_t, a_t)$, and 
 the expectation $\mathbf{E}$ is taken over user trajectories $\tau$ obtained by acting according to the policy: 
$a_t \sim \pi(\cdot | {s}_t)$, ${s}_{t+1} \sim \textbf{P}(\cdot | {s}_t, a_t)$. 

We build our method on top of the REINFORCE recommender introduced in \cite{williams1992simple}. Let the policy $\pi$ assume a functional form, mapping states to actions, parameterized by $\theta \in \mathbf{R}^d$. Using the log-trick, the gradient of the expected cumulative reward with respect to the policy parameters $\theta$ can be derived analytically \cite{williams1992simple}:
\begin{equation}
\label{eq:2}
      \nabla_{\theta}  \mathbf{E}_{\tau \sim \pi_{\theta}} [R(\tau)] =  \mathbf{E}_{\tau \sim \pi_{\theta}} \left[R(\tau) \nabla_{\theta}\log\pi_{\theta}(\tau)\right].
\end{equation}
To reduce variance in the gradient estimate a common practice is to discount the future reward with a discount $\gamma$: 
\begin{equation}
\label{eq:3}
   \sum_{\tau \sim \pi_{\theta}} [R(\tau) \nabla_{\theta}\log\pi_{\theta}(\tau)] \approx   \sum_{\tau \sim \pi_{\theta}} \sum_{t=0}^{|\tau|}  \left[R_t \nabla_{\theta}\log\pi_{\theta}(a_t|{s}_t)\right],
\end{equation}
where 
\begin{eqnarray}\label{eq:reward}
R_t =  & r({s}_t, a_t) + \gamma r({s}_{t+1}, a_{t+1}) + \gamma^2 r({s}_{t+2}, a_{t+2}) + \ldots  \nonumber\\
& + \gamma^{|\tau|-1-t} r({s}_{|\tau|-1}, a_{|\tau|-1}).
\end{eqnarray}

Equation \ref{eq:3} gives an unbiased estimate of the policy gradient in \emph{online RL}, where the gradient of the policy is computed on trajectories collected by the policy $\pi_{\theta}$ we are learning.
In practice, due to infrastructure limitations or production concerns, the trajectories available for learning are collected from a different \emph{logging} policy, or mixture of such policies, denoted by $\beta$ instead. Thus, we operate in an \emph{offline RL} setting, making the policy gradient as given by Eq.~\ref{eq:3} no longer unbiased. To address this skew, importance weighting is adopted~\cite{munos2016safe}. 
In this work, we also operate in batch offline RL, applying top-$K$ off-policy correction, and we defer readers to \cite{chen2019top} for details.



\section{Imputing Satisfaction in Reward}
\label{sec:proposed-approach}

We now turn to the main focus of this paper, i.e, shaping the reward of a REINFORCE recommender to drive user satisfaction. We start by describing how we parameterize the policy network (Section \ref{subsec:policy}); next we highlight the role of reward in REINFORCE for capturing long-term user utility  (Section \ref{subsec:reward}); and emphasize the challenges associated with considering satisfaction as a proxy to user utility (Section \ref{subsec:satisfaction-data}). Motivated by these challenges, we propose to augment the policy network with a satisfaction imputation network (Section \ref{subsec:actor-critic}).


\subsection{Policy Parameterization}
\label{subsec:policy}
We closely follow the setup in \cite{beutel2018latent,chen2019top} 
to parameterize the policy.
A Recurrent Neural Network (RNN) is used to encode the user's interaction history, capturing the changing user preferences. 
The output of the RNN is concatenated with the latent embeddings encoding context, which capture features like time of the day, device type. The concatenation of user sequential preferences and context embeddings is mapped to a lower dimensional representation via multiple Rectified Linear Units. This represents the user state $\mathbf{u}_s$. Conditioned on the user state $\mathbf{u}_s$, the policy $\pi_{\theta}(a|s)$ is then modeled
with a softmax,
\begin{equation}
    \pi_{\theta}(a|s) = \frac{\exp(\mathbf{u}_s^T \mathbf{v}_a /T)}{\sum_{a' \in \mathcal{A}} \exp(\mathbf{u}_s^T \mathbf{v}_{a'}/T)},
\end{equation}
where $\mathbf{v}_a$ are the action embeddings, and $T$ is a temperature term controlling the smoothness of the learned policy. 

\subsection{Reward}
\label{subsec:reward}
Reward plays a paramount role in determining the final learned policy. As shown in Equation~(\ref{eq:3}), the gradient from each state-action pair is weighted by the cumulative discounted reward $R_t$.
As prescribed in Equation~(\ref{eq:reward}), $R_t$ depends on the immediate rewards associated with the state-action pairs $r(s, a)$ as well as the discounting factor $\gamma$. In the absence of a user utility oracle, a key design choice is which \emph{proxy signals} to use to define the immediate reward. 
For each recommendation, the user could leave different signals indicating their experience with the item. Examples include implicit engagement-related signals, such as click, time spent engaging (reading/watching/listening), post-engagement actions, e.g., shares/likes/dislikes/comments, and they could leave explicit feedback in surveys asking them about their satisfaction level with the consumed item.





\begin{figure}[t]
    \centering
\centering
    \includegraphics[width=0.23\textwidth]{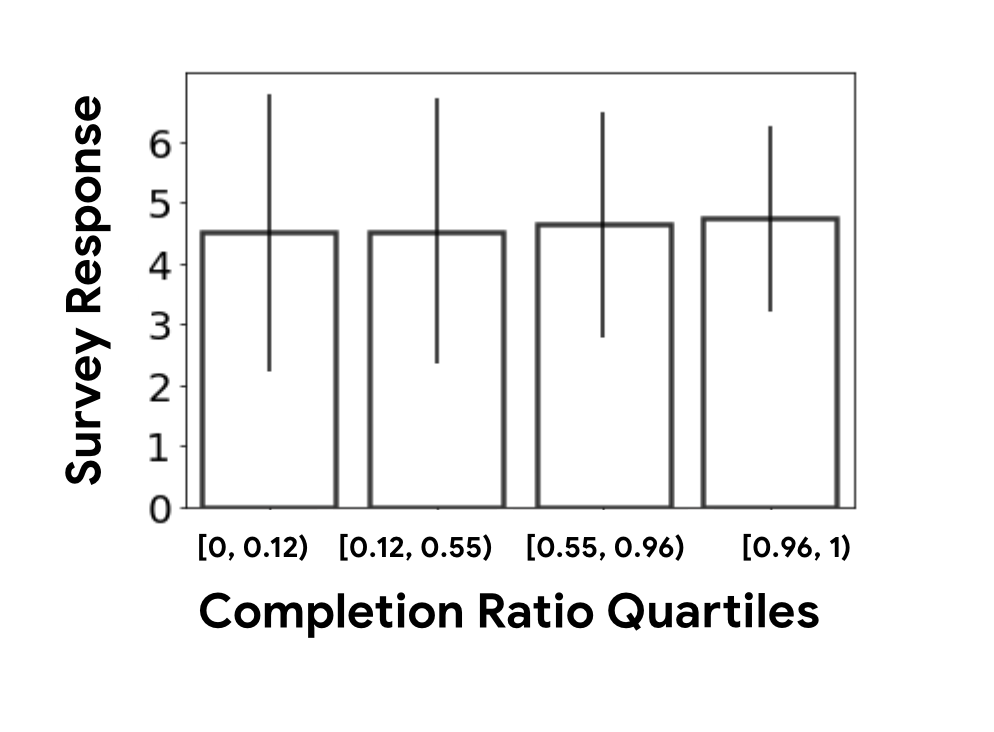}
    \includegraphics[width=0.23\textwidth]{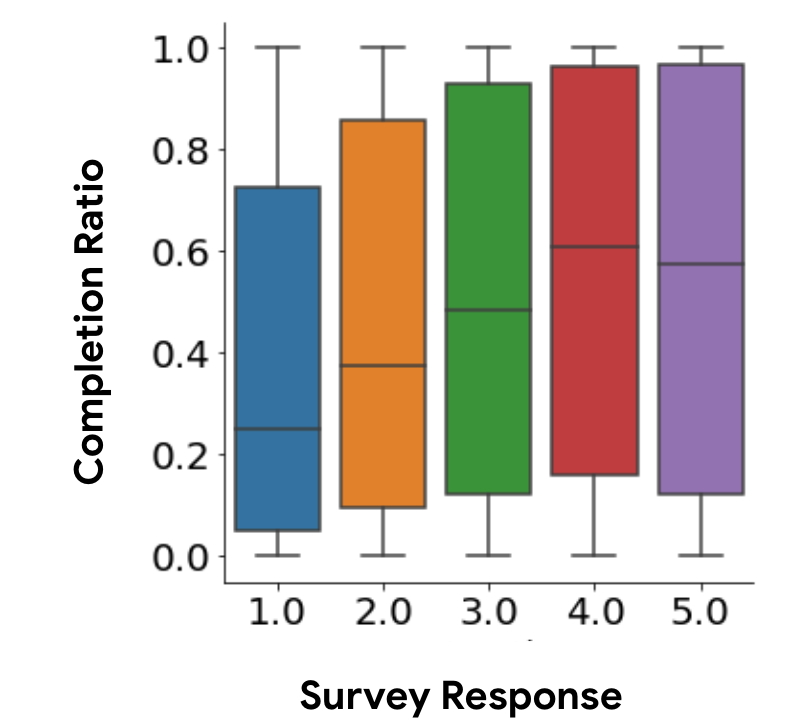}
    \label{fig:value_survey}
    \caption{Survey responses provide a  different set of information compared to behavior signals. Interactions with the same completion ratio can have vastly different associated satisfaction levels.}
    \label{fig:value_survey}
\end{figure}
\subsection{Value of Satisfaction Signals, and Challenges}
\label{subsec:satisfaction-data}
It is easy to see that if the
proxy signals used in the reward are solely engagement-focused, the policy will learn to choose actions that only drive engagement. This is not ideal as what users do (engagement) can be quite different from what they say they want (satisfaction), thus neglecting other important facets of the user experience. 

Figure \ref{fig:value_survey} illustrates this point. More than two million interactions with survey responses on a commercial recommendation platform were collected and analyzed. For the sake of this example, we consider completion ratio  (e.g., time spent on the item out of total length of the item) as one useful behavioral signal, and study its relationship with satisfaction signals as measured by survey responses in the scale of one to five. As shown in (Figure \ref{fig:value_survey} \emph{right}), grouping interactions by survey response rating, we find the higher the survey value, the higher the median completion ratio; however, we also see that per survey response value the range of associated completion ratios is quite large. This becomes more evident when grouping interactions with their associated survey responses based on the corresponding quantiles of completion ratios, (Figure \ref{fig:value_survey} \emph{left}). Based on the plotted 95\% confidence intervals,  interactions belonging in the exact same quantile of completion ratios (i.e., same user behavior), have quite different associated satisfaction levels  (Figure \ref{fig:value_survey} \emph{left}). 

It is worth pointing out that behavior signals alone fail to capture other sides of how the user felt about the interaction, e.g., did they find the content misleading, useful, did it provide some longer term value to them. It is therefore critical to consider both behavioral and satisfaction signals, and appropriately balance them when defining the reward.

Also, we opt for survey data rather than post-engagement signals as better proxies for user satisfaction as we have found that they can offer more representativeness---most users tend to not engage in post-click actions such as likes, dismissals.  Having said that, although we demonstrate the effectiveness of reward shaping with imputation networks for survey signals, the same technique is equally applicable for other proxy reward signals exhibiting similar concerns, such as likes, dislikes,  shares or dismissals. What is more, in our case study, we find that  satisfaction as measured by survey responses highly correlates with goodness as measured by post-engagement signals. Thus, we are able to significantly increase likes, and decrease dislikes/dismissals, even without explicitly optimizing for them (Section \ref{sec:live-experiments}).


If for each item the user interacted with in the trajectory, besides implicit engagement signals $r^e (s, a)$, we also had access to explicit satisfaction signals $r^u (s, a)$, we could define the immediate reward $r (s, a)$ as a function of the two, i.e.,
\begin{equation}
    r (s, a) = f\left(r^u (s, a), r^e (s, a)\right),
\end{equation}
where $f(\cdot)$ can include operators such as transformations on the raw signal (e.g., raising to a power, hinge, sigmoid) and combination functions (e.g., addition, multiplication) on the two reward signals. 

While the engagement signals $r^e$  are often dense, satisfaction signals $r^u$ are extremely sparse, as they are derived from user-provided  responses to satisfaction surveys. These surveys are shown uniformly to all users, asking them to rate on a scale how satisfying they found a sampled item from their recent engagement history. 
In our case study,  roughly \emph{one out of thousands of engagement signals will come with a satisfaction response}. This is because in a primarily content consumption-focused recommender platform, it would be disruptive to ask users to rate \emph{every} item consumed. Furthermore, users tend to not respond to surveys--- \emph{response rate is around 2\%} in our case. 

\subsection{Satisfaction Imputation Model}
\label{subsec:actor-critic}
This inherent sparsity of a subset of signals makes simply including them in the reward when present, ineffective. 

To address this challenge, we propose the use of an imputation network to densify the satisfaction signals, and include the imputed satisfaction signals in the reward instead. 

\begin{figure}[h]
    \centering
    \includegraphics[width=0.5\textwidth]{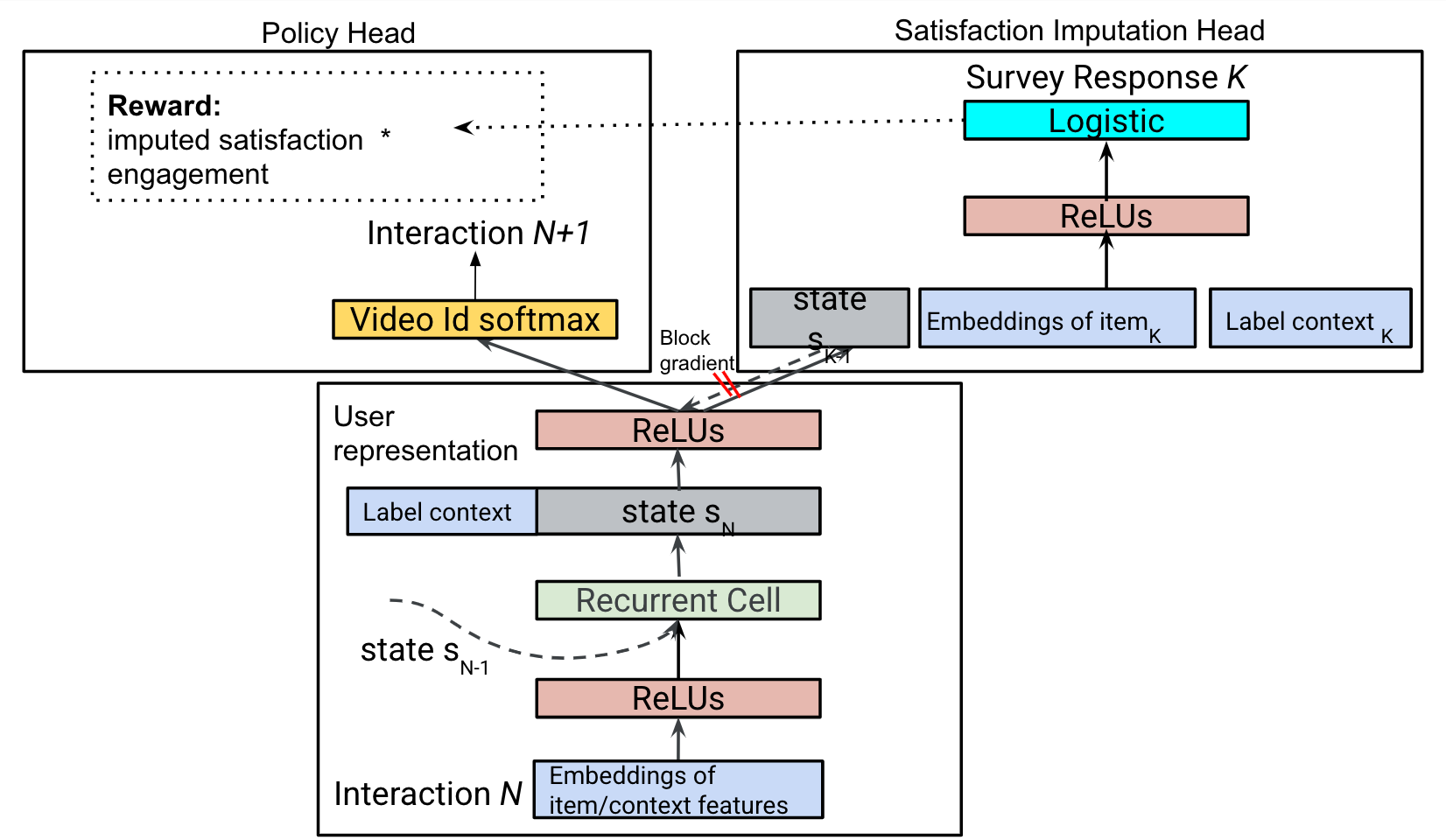}
    \caption{Proposed architecture, where a satisfaction imputation head is added to learn goodness of actions based on sparse survey responses, conditioned on state, action and context embeddings learned from the policy. The reward utilizes the satisfaction imputations to guide the policy head to select actions that lead to satisfied engagement.}
    \label{fig:actor-critic}
\end{figure}


The role of the imputation network is to map user state and action pairs $(s, a)$ to satisfaction scores, i.e., survey responses $sr$ present in the satisfaction data $\mathcal{D}_{sr}$. One can imagine learning a completely separate imputation model on this data, and then utilize these imputations directly on the reward of the policy network. The alternative which we opt for is to extend the policy network with a satisfaction imputation head and have  parameters shared between the two. This is quite appealing as 
given that the data used to train the policy head are of much higher volume compared to those used to train the imputation head, we hypothesize that transferring the learned user state and action embeddings from the dense task to the sparse one can be quite useful \cite{pan2009survey}.

Concretely, we propose a multi-task shared-bottom architecture with two heads, the policy head and the  satisfaction imputation head, each having their own task-specific parameters while sharing majorities of the state and action representations. As shown in Figure \ref{fig:actor-critic} (\emph{bottom}), the shared bottom encodes the sequential history of the user, as well as context information. The policy head in Figure \ref{fig:actor-critic} (\emph{upper left}) is identical to the standalone policy network described in Section \ref{subsec:policy}, with the only change being in its reward. 
The imputation head in Figure \ref{fig:actor-critic} (\emph{upper right}) is used to infer the satisfaction score for each state-action pair $(s, a)$ in the collected trajectories. Then, the imputed satisfaction score combined with engagement signals forms the reward used to train the policy network. 

We train the imputation head by gathering the corresponding state $\mathbf{u}_s$, and action embeddings $\mathbf{v}_a$ for any state-action pair associated with a survey response in the batch, and learning a mapping $\hat{sr}_\phi$ from these embeddings to the corresponding survey response i.e., $\hat{sr}_\phi: (\mathbf{u}_{s}, \mathbf{v}_{a}) \rightarrow sr,\forall sr \in \mathcal{D}_{sr}$. 
As shown in Figure \ref{fig:actor-critic} (upper right),  we prevent the imputation head from influencing the policy parameterization by stopping its gradient from flowing to these shared-bottom embeddings. 
To give the imputation head its own parameters to learn the mapping $\hat{sr}_\phi$, we concatenate the embeddings, i.e., $[\mathbf{u}_{s}, \mathbf{v}_{a}]$, and send them through multiple Rectified Linear Units (ReLU), and a final dense layer to map to the ground truth survey response. The ReLU layers and the dense layer  
are learned by optimizing an appropriate loss function $\ell$. For our case study, $\mathcal{D}_{sr}$  consists of survey responses in the scale of 1 to 5, with user studies showing that values of 4 and 5 are considered satisfying, whereas lower values show dissatisfaction. So, we considered a logistic loss, with a sigmoid for the last layer, to predict satisfying versus unsatisfying:
\begin{equation}\label{eq:critic}
    \min_{\phi} \sum_{sr \in \mathcal{D}_{sr}} \ell \left(\hat{sr}_\phi (\mathbf{u}_{s}, \mathbf{v}_{a}), sr\right).
\end{equation}

The policy head is learned via REINFORCE,
\begin{equation}
\label{eq:reinforce}
   \nabla_{\theta} \pi_{\theta} =  \sum_{\tau \sim \beta}  \sum_{t=0}^{|\tau|}  \left[ \frac{\pi_{\theta}(a_t | \mathbf{s}_t)}{\beta(a_t | \mathbf{s}_t)} \tilde{R}_t \nabla_{\theta}\log\pi_{\theta}(a_t|\mathbf{s}_t)\right]
\end{equation}
where $\tilde{R}_t$ denotes the imputed reward, and $\frac{\pi_{\theta}(a_t | \mathbf{s}_t)}{\beta(a_t | \mathbf{s}_t)}$ does the off-policy importance weighting. We decompose $\tilde{R}_t$ as
\begin{equation}
\label{eq:8}
    \tilde{R}_t = R_t^e \times \tilde{R}_t^u, 
\end{equation}
i.e., the ground truth engagement reward $R_t^e$, and the satisfaction reward $\tilde{R}_t^u$ predicted by the imputation network. 

The satisfaction imputation and the policy head are trained concurrently to optimize the (weighted) sum of the two losses.
In practice, to prevent a poorly estimated imputation head from corrupting the policy head, we start the training of the policy head with engagement only reward $R^e$, and include the imputed  $\tilde{R}^u$ only after the imputation head is properly trained. 
\begin{figure}[t!]
    \centering
    \includegraphics[width=0.25\textwidth]{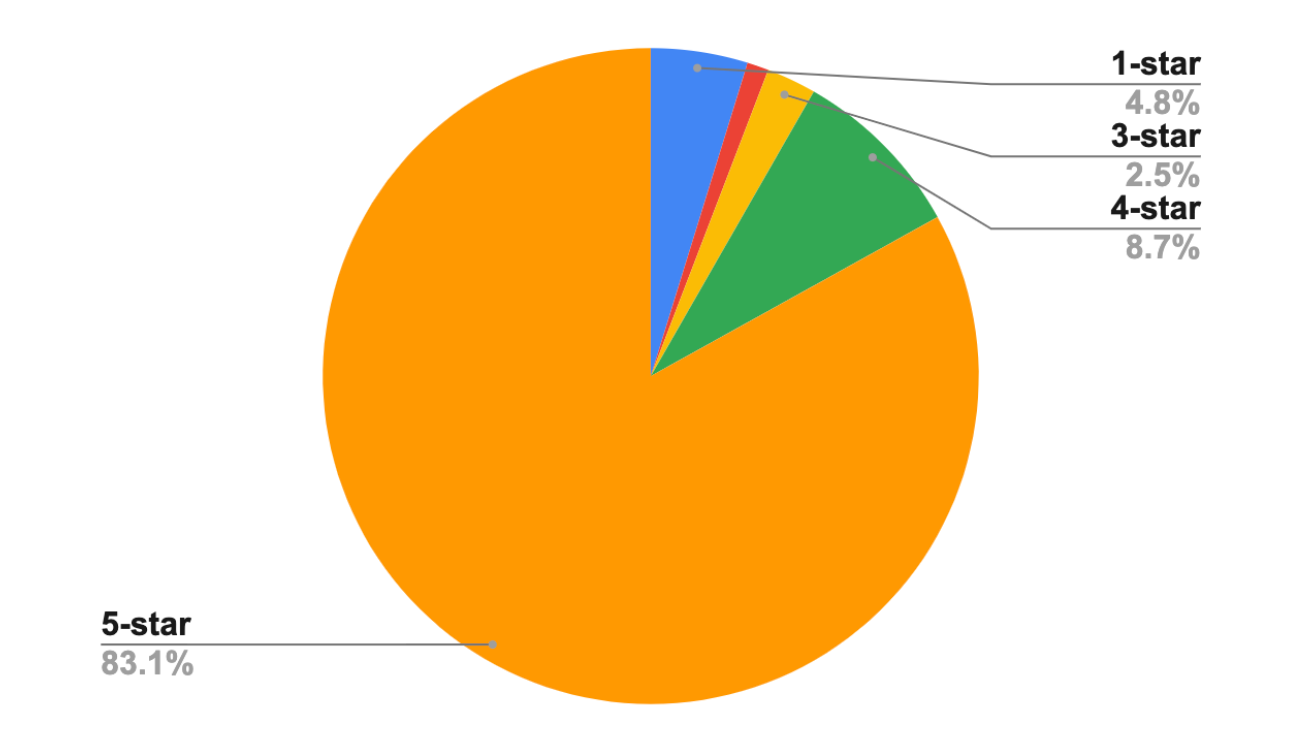}
    \includegraphics[width=0.22\textwidth]{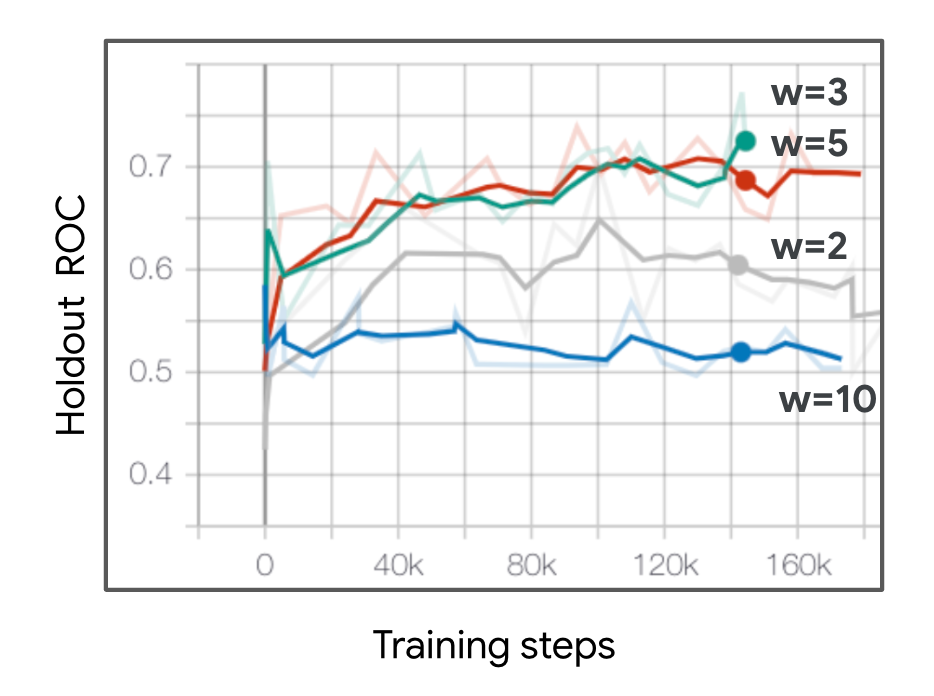}
    \caption{Addressing class imbalance (\emph{left}) by cost-sensitive learning (\emph{right}). Upweighting the negative class survey responses (1/2/3 star responses) can significantly improve the holdout satisfaction AUC ROC of the imputation model. However, increasing the weight too much can lead to a satisfaction accuracy deterioration, and a negative effect on the policy head's accuracy.}
    \label{fig:class-imbalance}
\end{figure}

\section{What makes a good Satisfaction Imputation Model?}
\label{sec:critic-results}

We now present some experimental findings on what makes a good satisfaction imputation model. To evaluate its predictive accuracy, we create a hold-out set consisting of user trajectories for users with at least one associated survey response. The AUC ROC achieved by the imputation model on the hold-out set is used as the offline evaluation metric. 

\textbf{Loss Function.} One challenge associated with the survey response data is the class imbalance problem. 
In our case study, the majority of responses recorded are in the higher spectrum (Figure \ref{fig:class-imbalance}, \emph{left}). One hypothesis is that users tend to respond to surveys about items they find highly satisfying ~\cite{paulhus1991measurement}. 
This creates a natural imbalance of survey values in the satisfaction data, leading the model to focus more on survey responses of higher values. An  under-specified model can predict every item to be satisfying as a result. 
One simple approach to address this is through cost-sensitive learning~\cite{elkan2001foundations}, in which the negative class of non-satisfying state-action pairs are weighted more. We calibrate the prediction after to reflect the ground-truth distribution of satisfying vs non-satisfying survey responses~\cite{chapelle2014simple}. 
Figure \ref{fig:class-imbalance} \emph{right} compares the performance of the satisfaction head with different weights on the negative class. We found a weight of 3 or 5 perform the best according to the holdout AUC ROC. 

\begin{figure}[t!]
    \centering
    \includegraphics[width=0.23\textwidth]{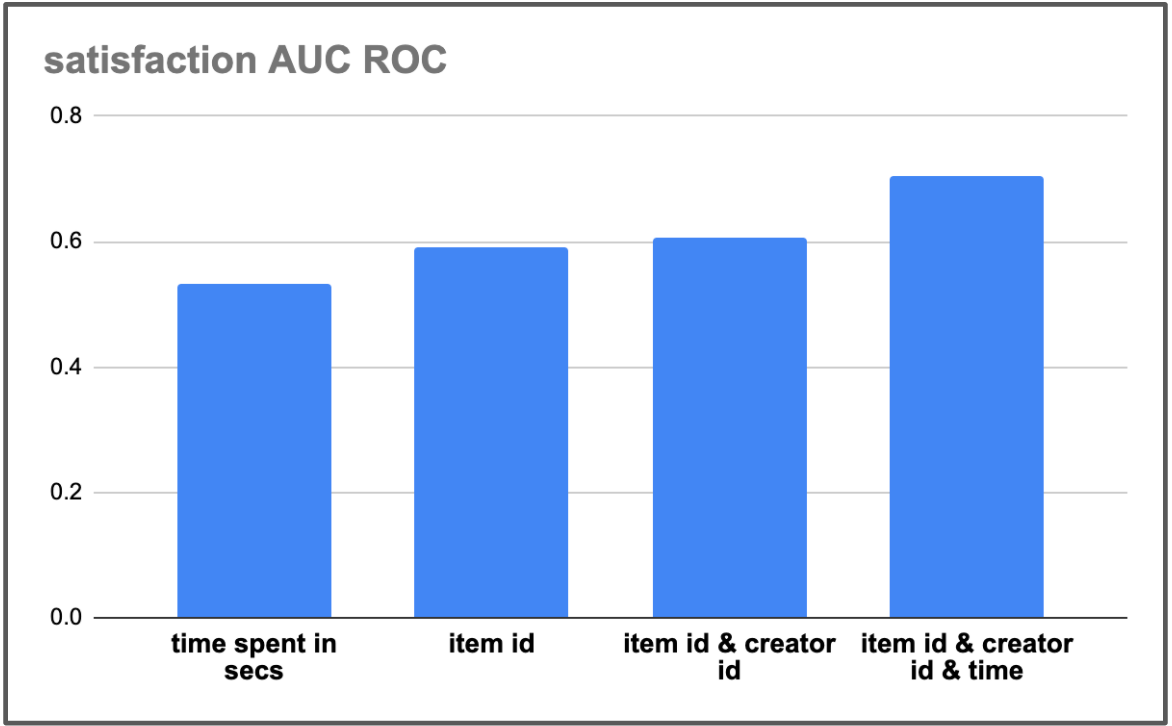}
    \includegraphics[width=0.23\textwidth]{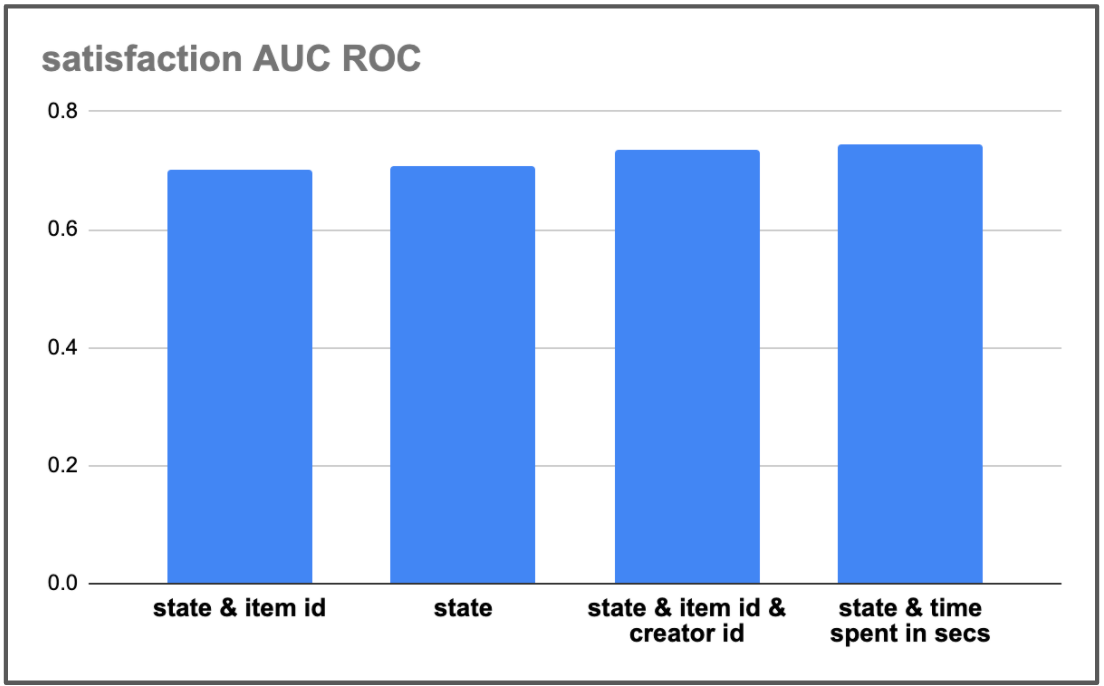}
    \caption{\emph{Left}: Using a set of features representing the item on the survey (i.e., item id \& creator id \& time spent on item in seconds), leads to an improved hold-out AUC ROC of the satisfaction imputation model, compared to using only individual action/item  features.~~ \emph{Right}: Using only user state (as outputted by the RNN, and concatenated with context embeddings) performs equally well with when including all action feature embeddings, without user state (as shown in \emph{top}). The satisfaction AUC ROC is further improved when including both user state and item-related features.}
    \label{fig:ROC_video_state}
\end{figure}

\textbf{Action Features.} 
Figure \ref{fig:ROC_video_state} \emph{left} summarizes the predictive power of different action features, i.e., time spent, item id and creator id, on the quality of the satisfaction imputation head. 
We can see that when using a single feature to predict survey response, the continuous feature of the time the user spent interacting with the item is less informative compared to discrete features representing the item such as the embedding of the item id.
The AUC ROC of a satisfaction imputation with item id as the only feature is further improved when including other features representing the item on the survey---we notice a slight improvement when including the creator id embedding, and a  considerable improvement when also including time spent in seconds interacting with the item, on top of item id and creator id embeddings. 


\textbf{User State.}
We also evaluate the importance of including user state in learning the imputation model (Figure \ref{fig:ROC_video_state} \emph{right}). We can see that when using as features only the user state, as captured by the RNN over the sequence up until this point, concatenated with the label context embedding, we get the same hold out AUC ROC as the one achieved by using all features representing the action (Figure \ref{fig:ROC_video_state} \emph{left}, last bar). Concatenating the user state with action embeddings further improves the imputation model's predictive power.  Thus, in what follows, our satisfaction imputation model will utilize both user state and action embeddings as features.


\section{Live Experiments}
\label{sec:live-experiments}


Our case study involves a large scale two-stage recommender platform, where at the first stage multiple candidate generators retrieve potential candidates from the entire corpus; and the second stage involves a ranker model ranking the candidates and providing a final top-$K$ recommendation list to be shown to the user.

To study the extent to which  our approach can improve real user experiences, we apply our reward shaping approach onto a RL-based candidate generator, and conducted a series of A/B experiments. 
The control arm runs a REINFORCE agent learned using engagement-only reward. 
In the experiment arm, we test our proposed approach of augmenting the policy network with a satisfaction imputation head (Figure \ref{fig:actor-critic}), and utilizing the imputed satisfaction reward along with the ground truth behavioral signals into the policy's reward,  described in \ref{subsec:actor-critic}. 

Experiments are run for over a month on a fixed set of randomly assigned user traffic to study the long-term effect. During this period, the model is trained continuously,  with new interactions being used as training data with a lag under 24 hours. 


\textbf{Online Satisfied Engagement Metric.} For evaluating whether the user experiences are improved, one could look at ground truth survey responses. An experiment which increases the average survey response over the experiment period would be considered driving more user satisfaction. In fact our experiment increases 5 star survey responses on average by \textbf{0.48\%} and decreases 1 and 2 star survey responses by \textbf{1.89\%}. However, we again run into the key challenge  associated with survey responses which is  sparsity. If we only measure on user-item pairs for which the users have responded to in a survey, we would only be looking at a very small percentage of the user interactions.  
To tackle this, 
we instead rely on a model-based metric predicting a survey response for each of the items the user has interacted with, and combining that with engagement metrics measured live. It is worth pointing out, the model used for measuring online satisfaction metric is independent of the imputation network we built, with a considerably different feature set and architecture. We cannot utilize the predictions of this model directly into our reward, due to infrastructure complexities and freshness requirements. 

 \begin{figure}[!t]
    \centering
    \includegraphics[width=0.4\textwidth]{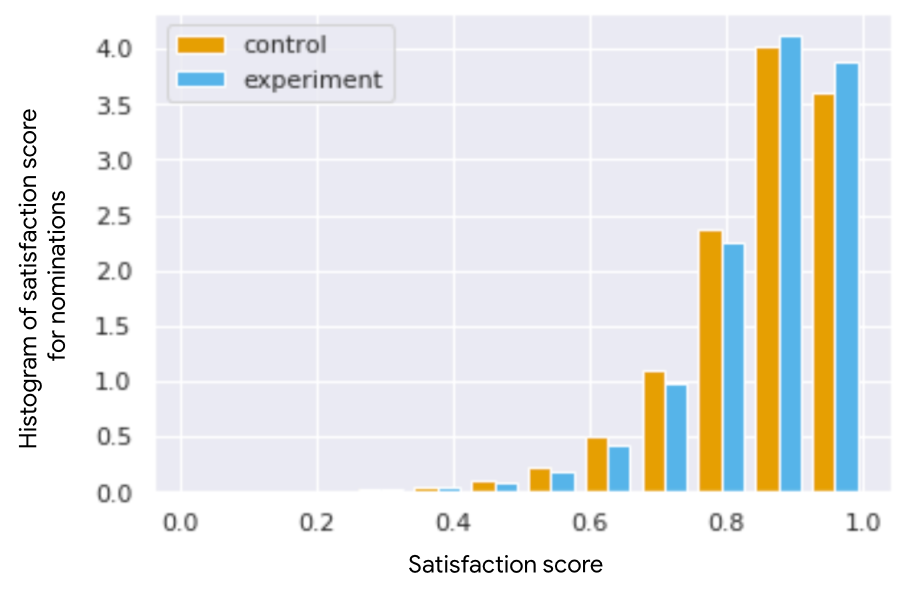}
    \caption{Distribution of satisfaction scores for items nominated by model, for experiment arm (REINFORCE with satisfaction imputation model) versus control arm (REINFORCE without satisfaction in reward).}
    \label{fig:barnett_distribution}
\end{figure}
\textbf{Satisfaction Improvements.} In Figure \ref{fig:barnett_distribution} we show how the distribution of ground truth satisfaction scores over nominations coming from the control model (REINFORCE with engagement-reward) versus the experiment (our approach optimizing for a combination of satisfaction and engagement) compare. The x-axis represents the ground truth satisfaction probability score (used to calculate the live satisfied engagement time metric, and distinct from our imputation model predictions), with scores close to 1.0 indicating users being satisfied with their interactions, and close to 0.0 being unsatisfied. We can see that in both experiment and control arms, the majority of interactions are predicted to have a score larger than 0.5, indicating satisfying experience. Nevertheless, we can clearly see that our experiment increases the number of nominations with satisfaction scores greater or equal to 0.9, and decreases respectively nominations with a score less than 0.9. This demonstrates that our proposed imputation head is able to identify items which are satisfying to the users and shift the policy to select  more satisfying items, further validating its predictive accuracy. 

\begin{figure}[!t]
    \centering
    \includegraphics[width=0.4\textwidth]{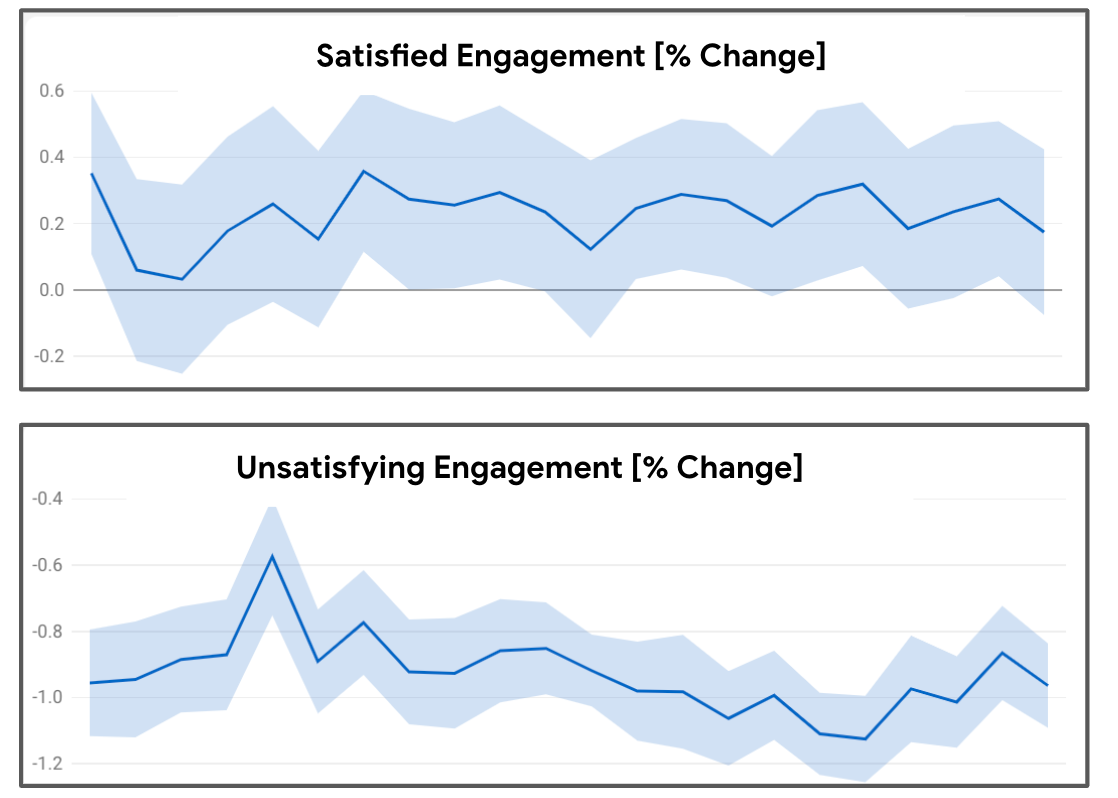}
    \caption{Percentage improvements of online satisfaction metrics (y-axis) achieved by our proposed model over the course of a month (x-axis). Satisfied engagement is significantly increased by 0.23\% (\emph{top}), while unsatisfying engagement 
is significantly decreased by 0.93\% (\emph{bottom}), highlighting the value of reward shaping for driving user satisfaction.}
    \label{fig:overall-live}
\end{figure}

Figure \ref{fig:overall-live} compares the control and experiment arm on a live metric combining the model-based satisfaction metric and behavioral-based implicit engagement signals. 
We find that on average, satisfied engagement is increased by \textbf{0.23\%}, while unsatisfying engagement 
is decreased by \textbf{0.93\%}. Both results are statistically significant, signifying the value of reward shaping in driving user utility as specified by the reward. 

Furthermore, metrics orthogonal to the ground truth satisfaction scores, measuring other facets of user experience, significantly move towards the right direction: likes on items increase by \textbf{0.53\%}, while dislikes are decreased by \textbf{1.11\%} and dismissals decrease by \textbf{3.03\%}. Note that we did not include these signals into the features or labels of our satisfaction imputation model. This further supports the point made in Section \ref{sec:intro} that optimizing for satisfaction signals correlates well with improvements in post-engagement actions.

\begin{figure}[t!]
    \centering
    \includegraphics[width=0.4\textwidth]{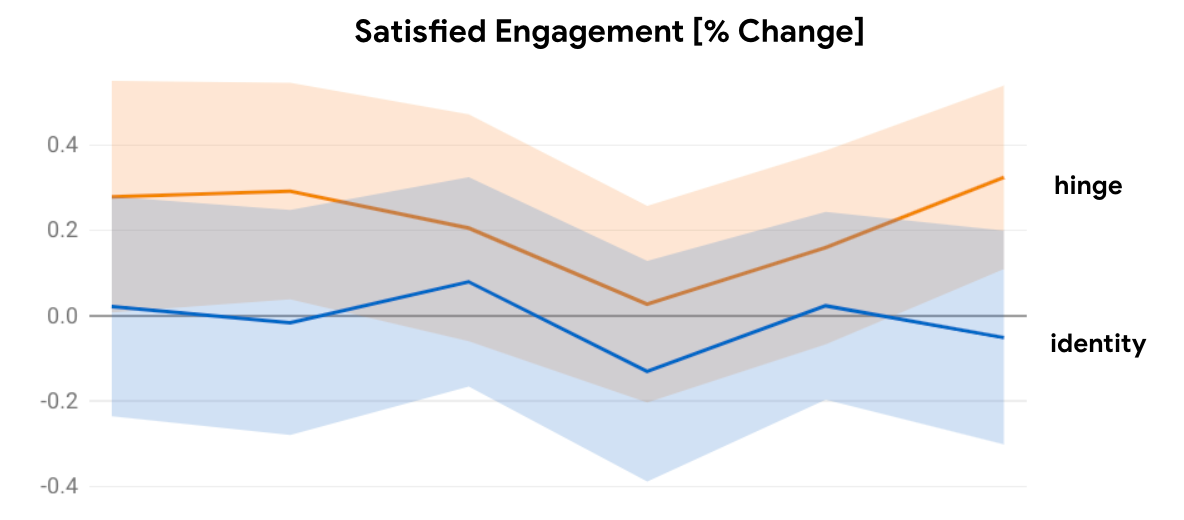}
    \caption{Effect of choice of transformation function over the imputed satisfaction scores. Y axis denotes percentage improvements of online satisfied engagement metric,  x-axis denotes the days over the course of the month, and the two lines refer to hinge and identity transformation. We find that filtering data with lower imputed satisfaction scores through the hinge function performed considerably better compared to raw imputations.}
    \label{fig:live_hinge_identity}
\end{figure}

We found that an important design choice in the reward  
is the transformation function over the predicted satisfaction signal $r^u(s, a)$ by the imputation network, i.e., the probability of an item being satisfying to the user. Simply multiplying the engagement reward signals by the imputed probability of the item being satisfying (identity function) only decreases non-satisfied engagement, but did not lead to statistically significant improvements in satisfied engagement. It is critical to further differentiate highly satisfying items from less satisfying ones to allow the model to clearly prefer selecting such items. We found that in practice a simple hinge function  performed the best, i.e., when imputed probability of an item being satisfying to the user is larger than a threshold, multiply with the probability, else completely zero out the engagement reward (Figure \ref{fig:live_hinge_identity}). The threshold was tuned offline based on the ground truth response distribution, and the imputation network's predictions. We report results based on threshold set to 0.75. 

Furthermore, we validated in live experiments some of our choices made offline. We found that predicting the probability of an item found satisfying by the user performed better than predicting the actual survey response they will give, i.e., cross-entropy loss gave better results compared to a square loss. We hypothesize that this could be the case due to better alignment with the loss used to train the model-based Satisfied Engagement live metric. Also, balancing the data to give more weight to unsatisfying survey responses and calibrating the prediction to account for the balancing (Figure \ref{fig:class-imbalance}) was important for live improvements. Finally, we found that raising the satisfaction reward term to an exponent larger than 1 gave us slightly better results when using the identity transformation function; but for the hinge function, the improvement was not statistically significant.




\section{Conclusions}
\label{sec:conclusions}
In this paper, we considered the problem of driving long-term user satisfaction in a reinforcement learning  recommender. We posited that reward shaping is a powerful tool for aligning the RL recommender's objective with what users want. We argued that engagement signals only, as typically considered by existing RL literature, are not able to capture the full aspects of user experience on the recommendation platform. 
Instead, satisfaction signals, capturing how users felt about items they have interacted with, should be incorporated into the reward. We highlighted that a key challenge associated with such signals, when trying to incorporate them into the reward, is their sparsity---only a small percentage of the user-item interactions has associated survey response value. To combat the sparsity, we proposed to extend a state-of-the-art REINFORCE recommender with a satisfaction imputation network, imputing for every interacted item the user's satisfaction score. We offered insights based on offline model improvements, and demonstrated via live experiments in a commercial large-scale recommender that including satisfaction imputation into the reward indeed drives more satisfying user experiences.

\bibliography{example_paper}
\bibliographystyle{icml2021}

\end{document}